\documentclass[aps,%prd,
superscriptaddress,twoside,%twocolumn,
nofootinbib,preprintnumbers]{revtex4}

\usepackage{amsmath}
\usepackage{graphicx}

\newcommand{\lesssim}{\mathrel{\mathpalette\vereq<}}
\newcommand{\gtrsim}{\mathrel{\mathpalette\vereq>}}

\begin{document}
\title{Note on anomalous dimension in  QCD
}      
\author{{Koichi Yamawaki}} \thanks{
      {\tt yamawaki@kmi.nagoya-u.ac.jp}}
      \affiliation{ Kobayashi-Maskawa Institute for the Origin of Particles and the Universe (KMI) \\ 
 Nagoya University, Nagoya 464-8602, Japan.}
 \begin{abstract} 
The anomalous dimension 
$\gamma_m =1$ in the infrared region  near conformal edge in the broken phase of the large $N_f$ QCD  has been shown by the ladder Schwinger-Dyson equation 
and also by the lattice simulation for $N_f=8$ for $  N_c=3$. 
Recently Zwicky  claimed  another independent 
 argument (without referring to explicit dynamics) for the same result,  $\gamma_m =1$,  by comparing 
$\left<\pi(p_2)|(1+\gamma_m)\cdot   \sum^{N_f}_{i=1} m_f \bar \psi_i \psi_i |\pi(p_1)\right>= \left<\pi(p_2)|\theta_\mu^\mu|\pi(p_1)\right>=2 M_\pi^2$ (up to trace anomaly)  with his estimate  of $\left<\pi(p_2)| 2\cdot  \sum^{N_f}_{i=1} m_f \bar \psi_i \psi_i |\pi(p_1)\right> =2 M_\pi^2$ 
 through  Feynman-Hellmann theorem combined  with an assumption 
 $M_\pi^2\sim m_f$  
 characteristic to the broken phase. 
We show that this is not justified by explicit evaluation of each matrix element based on the  
``dilaton chiral perturbation theory  (dChPT)'' : 
 $<\pi(p_2)| 2\cdot  \sum^{N_f}_{i=1} m_f \bar \psi_i \psi_i |\pi(p_1)>= 
 2M_\pi^2 + [(1-\gamma_m) M_\pi^2\cdot 2/(1+\gamma_m)]= 2 M_\pi^2 \cdot 2/(1+\gamma_m)
 \ne 2 M_\pi^2$
  in contradiction with his estimate,  which is compared with
 $<\pi(p_2)| (1+\gamma_m) \cdot  \sum^{N_f}_{i=1}  m_f \bar \psi_i \psi_i |\pi(p_1)> =(1+\gamma_m) M_\pi^2+ [(1-\gamma_m) M_\pi^2]=2 M_\pi^2$ (both up to trace anomaly),  
where the  terms in $[ \,\,]$ are  from the $\sigma$ (pseudo-dilaton) pole contribution. Thus there is no constraint on $\gamma_m$ when the $\sigma$ pole 
 contribution is  treated consistently for both. We further show that Feynman-Hellmann theorem  is applied to the inside of the conformal window where dChPT is invalid and the $\sigma$ pole contribution is absent, 
 with $M_\pi^2 \sim m_f^{2/(1+\gamma_m)}$ instead of  $M_\pi^2 \sim m_f$, we have the same result as ours in the broken phase. 
Further  comment related to dChPT is made on the decay width of  $f_0(500)$ to $\pi\pi$ for $N_f=2$.  
It  is shown to be consistent with the reality, when {\it both $\pi$ and $f_0(500)$ are regarded as pseudo-NG bosons  with the non-perturbative trace anomaly dominance}. 
\end{abstract}
\maketitle

\section{Introduction}
The anomalous dimension 
$\gamma_m =1$  together with the  pseudo-dilaton in the  gauge theory with the spontaneously broken chiral/scale symmetry is an essence of the walking technicolor \cite{Yamawaki:1985zg,Bando:1986bg}.   
 It has been shown by the ladder Schwinger-Dyson equation~\cite{Matsuzaki:2015sya} and also by the lattice simulation for QCD with $N_f=8$ through (approximate) hyperscaling fit~\cite{LatKMI:2016xxi}.
Recently Roman Zwicky~\cite{Zwicky:2023bzk}
 claimed  another independent argument (without referring to explicit dynamics) for the same result,  $\gamma_m =1$, near conformal edge in the broken phase of the large $N_f$ QCD:
He evaluated 
 \begin{eqnarray}
 <\pi(p_2)| 2\cdot  \sum^{N_f}_{i=1} m_f \bar \psi_i \psi_i |\pi(p_1)> =2 M_\pi^2 
 \end{eqnarray}  
 through the Feynman-Hellmann theorem combined with an additional assumption $M_\pi^2 \sim m_f$.  He further showed that the result coincides  with 
 the double use of the soft pion theorem. This was then compared with the standard generic 
evaluation of the matrix element 
$\left<\pi(p_2)|(1+\gamma_m)\cdot   \sum^{N_f}_{i=1} m_f \bar \psi_i \psi_i |\pi(p_1)\right>=\left<\pi(p_2)|\beta(\alpha)/(4 \alpha)\cdot G_{\mu\nu}^2 + (1+\gamma_m)  \sum^{N_f}_{i=1} m_f \bar \psi_i \psi_i |\pi(p_1)\right>
= \left<\pi(p_2)|\theta_\mu^\mu|\pi(p_1)\right>=2 M_\pi^2$, 
with an additional assumption of the IR fixed point (even in the broken phase and $M_\pi^2 \ne0$) to ignore the trace anomaly contribution. Then he concluded  $\left<\pi(p_2)|(1+\gamma_m) \cdot  \sum^{N_f}_{i=1}m_f  \bar \psi_i \psi_i |\pi(p_1)\right>=\left<\pi(p_2)| 2\cdot    \sum^{N_f}_{i=1} m_f \bar \psi_i \psi_i |\pi(p_1)\right> $, i.e.,
 $\gamma_m=1$ ($=\gamma_*$ under his assumption of IR fixed point with vanishing trace anomaly).

In this paper,  we show that explicit evaluation of each matrix element based on the nonlinear realization Lagrangian of scale and chiral symmetries, ``dilaton chiral perturbation theory  (dChPT)''~\cite{Matsuzaki:2013eva}\cite{Leung:1989hw}  gives 
 \begin{eqnarray}
 <\pi(p_2)| 2\cdot  \sum^{N_f}_{i=1} m_f \bar \psi_i \psi_i |\pi(p_1)> &=&2M_\pi^2 + \left[\frac{2}{1+\gamma_m} \cdot (1-\gamma_m) M_\pi^2\right]= \frac{2}{1+\gamma_m}\cdot  2 M_\pi^2,
 \label{dChPTresult1} 
 \end{eqnarray}
 \begin{eqnarray}
 <\pi(p_2)|(1+\gamma_m)\cdot   \sum^{N_f}_{i=1} m_f \bar \psi_i \psi_i |\pi(p_1) > &=&(1+\gamma_m) M_\pi^2 + \left[(1-\gamma_m) M_\pi^2 \right]= 2 M_\pi^2,
  \label{dChPTresult2}
    \end{eqnarray}
(both up to trace anomaly), where the terms in $[ \quad ]$ are from the $\sigma$ pole contribution. Note that Eq.(\ref{dChPTresult2}) is consistent with the well-known generic result $ <\pi(p_2)|\theta_\mu^\mu|\pi(p_1) >=2 M_\pi^2$  
based on the form factor argument only when including the $\sigma$ pole contribution. 
 Thus including (or ignoring) the $\sigma$ pole contribution {\it for both matrix element consistently},   there is no constraint on $\gamma_m$ in contrast to the Zwicky's result. 
 Even including the trace anomaly, we will show that the result keeps the relation $\left<\pi(p_2)| (1+\gamma_m) \cdot  \sum^{N_f}_{i=1}  m_f \bar \psi_i \psi_i |\pi(p_1)\right> =(1+\gamma_m)/2 \cdot  \left<\pi(p_2)| 2\cdot  \sum^{N_f}_{i=1} m_f \bar \psi_i \psi_i |\pi(p_1)\right>$,
consistent with Eqs.(\ref{dChPTresult1}) and (\ref{dChPTresult2}).

 The same result is also obtained 
 based on the Feynman-Hellmann theorem, within the conformal window where dChPT is invalid and no $\sigma$ pole contribution exists,  with $\pi$ now  as a non-pseudo NG boson having the mass
 to obey the hyperscaling, $M_\pi^2 \sim m_f^{2/(1+\gamma_m)}$,  instead of the pseudo-NG boson case $M_\pi^2 \sim m_f$ in the broken phase:~\footnote{ He unjustifiably identifies  this as the hyperscaling with $\gamma_m=1$ in the 
 generic broken phase (including the deeply broken phase like $N_f=2$). It was shown on the lattice~\cite{LatKMI:2016xxi} that for $N_f=4$  generic hadron spectra (including $F_\pi$) other than $M_\pi$ do not obey the hyperscaling at all
 and hence  $M_\pi^2 \sim m_f$ cannot be understood as hyperscaling. For $N_f=8$ near the conformal window,  on the other hand, spectra other than $M_\pi$  do obey the hyperscaling with $\gamma_m \simeq 1$, while $M_\pi$ does only non-universally with $\gamma_m\sim 0.6$ due to $m_f$ dependence away from the chiral limit as a pseudo NG boson  is different from the others obeying hyperscaling.  }
 \begin{eqnarray}
  <\pi(p_2)| 2\cdot  \sum^{N_f}_{i=1} m_f \bar \psi_i \psi_i |\pi(p_1)> &=&2  \frac{\partial}{\partial \ln m_f}  <\pi(p_2)| {\cal H} |\pi(p_1)>  = 
  \frac{\partial}{\partial \ln m_f} 2 E_\pi^2 \nonumber \\
 &=& \frac{2}{1+\gamma_m} 2 M_\pi^2 \ne 2 M_\pi^2, \nonumber\\
  <\pi(p_2)| (1+\gamma_m) \cdot  \sum^{N_f}_{i=1} m_f \bar \psi_i \psi_i |\pi(p_1)> &=&(1+\gamma_m)  \frac{\partial}{\partial \ln m_f}   <\pi(p_2)| {\cal H} |\pi(p_1)> 
  = \frac{\partial}{\partial \ln m_f} (1+\gamma_m)  E_\pi^2\nonumber\\
  &=&  \frac{2}{1+\gamma_m} (1+\gamma_m) M_\pi^2= 2 M_\pi^2,
      \end{eqnarray}
the same as Eqs.(\ref{dChPTresult1}) and (\ref{dChPTresult2}) in the broken phase (up to the trace anomaly term). Thus the result is independent of the phases, broken or conformal, as it should be.  Actually, 
 the Feynman-Hellmann theorem is insensitive to the spontaneous symmetry breaking, giving the same kinetic term form in $M_\pi^2$ independently of the phase, and the combined use of
 $M_\pi^2 \sim m_f$ characteristic to the broken phase is not justified, which would result in $  <\pi(p_2)| 2\cdot  \sum^{N_f}_{i=1} m_f \bar \psi_i \psi_i |\pi(p_1)> = 2M_\pi^2$ and 
 $  <\pi(p_2)| (1+\gamma_m) \cdot  \sum^{N_f}_{i=1} m_f \bar \psi_i \psi_i |\pi(p_1)>   = (1+\gamma_m) M_\pi^2$, the same as the wrong results neglecting the $\sigma$ pole contribution in Eqs.(\ref{dChPTresult1}) and (\ref{dChPTresult2}). 
 If the theorem were to be used in the broken phase, then  all the hadron masses including $M_\pi$ 
 should be regarded as a simple Coulombic bound state  $M_H \sim  2 m_f^{(R)} \sim m_f^{1/(1+\gamma_m)}$ as in the 
 conformal phase, in which case the result would coincide with the correct one.
 
As to the double soft pion theorem for $<\pi(p_2)| 2\cdot  \sum^{N_f}_{i=1} m_f \bar \psi_i \psi_i |\pi(p_1)>$  which he claims gives equivalent result as that from the Feynman-Hellmann theorem combined with his assumption $M_\pi^2 \sim m_f$,  it ignores the $\sigma$ pole contribution, the term in  $[ \quad ]$ of  Eq.(\ref{dChPTresult1}).
Actually, the same double soft pion theorem applied consistently for both matrix element would give  $\left<\pi(p_2)| 2 \cdot  \sum^{N_f}_{i=1}m_f  \bar \psi_i \psi_i |\pi(p_1)\right> =2 M_\pi^2$
and $\left<\pi(p_2)| (1+\gamma_m) \cdot  \sum^{N_f}_{i=1}  m_f \bar \psi_i \psi_i |\pi(p_1)\right> =(1+ \gamma_m) M_\pi^2$, 
thus again no constraint on the value of $\gamma_m$ (or $\gamma_*$). 
Inclusion of the $\sigma$ pole contribution for both 
 gives the correct 
 results (up to the trace anomaly):  $\left<\pi(p_2)| 2\cdot  \sum^{N_f}_{i=1} m_f \bar \psi_i \psi_i |\pi(p_1)\right>= 
 2M_\pi^2 + [(1-\gamma_m) M_\pi^2\cdot 2/(1+\gamma_m)]= 2 M_\pi^2 \cdot 2/(1+\gamma_m)\ne 2 M_\pi^2$,  while
 $\left<\pi(p_2)| (1+\gamma_m) \cdot  \sum^{N_f}_{i=1}  m_f \bar \psi_i \psi_i |\pi(p_1)\right> =(1+\gamma_m) M_\pi^2+ [(1-\gamma_m) M_\pi^2]=2 M_\pi^2$ to be consistent with the generic form factor argument, 
  where the  term in $[ \,\,]$ of each result is from the $\sigma$ (pseudo-dilaton) pole contribution. Thus  there is no constraint on $\gamma_m$ when the $\sigma$ pole 
 contribution is consistently included/ignored for both.

Also we shall make a comment related to dChPT on the decay width of  $f_0(500)$ to $\pi\pi$ for $N_f=2$ where the scale symmetry spontaneously broken is also broken  explicitly by the non-perturbative trace anomaly and
the quark mass. It  is shown to be consistent with the reality, when {\it both $\pi$ and $f_0(500)$ are regarded as  pseudo-NG bosons}, based on this dChPT, with the non-perturbative trace anomaly dominance.  This is contrasted with 
the decay width evaluated by
 the
low energy theorem for the scale symmetry, which regards $f_0(500)$ as a pseudo NG boson but $\pi$ as a matter field not as pseudo NG boson and gives 50 times smaller than the real data, long-standing problem  
and has  long been a puzzle 
when $f_0(500)$ is regarded as a pseudo-dilaton $\sigma$.

\section{Nonlinear realization of the chiral and scale symmetries}

Let us start with the basic formula based on the Ward-Takahashi (WT) identity for $N_f$ QCD (with the same mass $m_f$ for $N_f$ flavors) for $\theta_\mu^\mu$~\cite{Matsuzaki:2015sya}:
\begin{eqnarray}
\theta^\mu_\mu=\partial_\mu D^\mu=\frac{\beta^{(\rm NP)}(\alpha)}{4\alpha} G_{\mu\nu}^2+
 (1+\gamma_m)   \sum^{N_f}_{i=1} m_f \bar \psi_i \psi_i ,
\label{thetamumu}
\end{eqnarray}
with $\psi_i$ for a single flavor within the  degenerate $N_f$ flavors,  and $\frac{\beta^{(\rm NP)}}{4\alpha} G_{\mu\nu}^2$ is 
the {\it non-perturbative} trace anomaly, $\left<0|\frac{\beta^{(\rm NP)}(\alpha)}{4\alpha} G_{\mu\nu}^2|0\right> =-{\cal O}(\Lambda_{\rm IR}^4)$ (up to factor $N_f N_c$),
due to the dynamically generated  IR mass scale $\Lambda_{\rm IR}$ (or dynamical quark mass $m_D\sim M_\rho/2\sim M_N/3$)
 in the chirally broken phase with $\left<0|(\bar \psi \psi)_R |0)\right> =- {\cal O} (\Lambda_{\rm IR}^3)$. 
 Here the perturbative trace anomaly $<\frac{\beta^{(\rm perturbative)}}{4\alpha} G_{\mu\nu}^2> =- {\cal O} (\Lambda_{\rm QCD}^4)$ due to the regularization, with 
 the UV scale $\Lambda_{\rm QCD}$ characterizing the asymptotically-free running of the perturbative coupling,  is irrelevant to the IR physics thus subtracted out from Eq.(\ref{thetamumu}). 
 \footnote{In the broken phase near the conformal window with $\alpha_* \gtrsim \alpha_{\rm cr}$, where $\alpha_{\rm cr}$  is the critical coupling for the condensate to be generated. The dynamically generated fermion mass $m_D$ takes the  form of the 
 essential singularity, Miransky-BKT (Berezinsky-Kosterlitz-Thouless)  type: $m_D \sim \Lambda \cdot \exp [-a/(\alpha -\alpha_{\rm cr})^r] 
 \rightarrow 0$ \, $(a, r>0)$,  for 
 $\alpha (\lesssim \alpha_*) \searrow \alpha_{\rm cr}$, where $\Lambda$ is the UV scale to be identified with the intrinsic scale 
$\Lambda_{\rm QCD}$. {\it Due to $m_D\ne 0$   the perturbative IR fixed point $\alpha_*$ is washed out in contradiction with the Zwicky's assumption. }
  The coupling for $\alpha>\alpha_{\rm cr}$  runs non-perturbatively due to this mass generation,
  with  $\beta^{(\rm NP)}(\alpha)$ having now  a UV fixed point at $\alpha_{\rm cr}$ instead of IR fixed point:
The ladder SD equation gives $a=\pi, r=1/2, \alpha_{\rm cr}=\pi/4$ for $N_c=3$ near the conformal window $\alpha_*  \gtrsim \alpha_{\rm cr}$ and
 $\beta^{({\rm NP})}(\alpha)= \frac{\partial \alpha(\Lambda)}{\partial \ln \Lambda}$
$=-\frac{2 \alpha_{\rm cr}} {\pi} \left(\frac{\alpha}{\alpha_{\rm cr}} -1\right)^{3/2}$ 
which vanishes at $\alpha\searrow \alpha_{\rm cr}$, 
while $\left<0|G_{\mu\nu}^2|0\right> 
\sim  \left(\frac{\alpha}{\alpha_{\rm cr}} -1\right)^{- 3/2} m_D^4$ 
blows up, to precisely cancel the vanishing $\beta^{({\rm NP})}(\alpha)$, resulting in $<0|\frac{\beta^{(\rm NP)}(\alpha)}{4\alpha} G_{\mu\nu}^2|0>=- {\cal O} (m_D^4)$. See, e.g., Ref.\cite{Matsuzaki:2015sya}.
\label {fixedpoint}
}
  
 From the  pole-dominated  WT identity for Eq.(\ref{thetamumu}) we have:
   \begin{eqnarray}
  M_\sigma^2 F_\sigma^2&= i  {\cal F.T.}
 \left.\langle 0| T
  \left(\partial_\mu D^\mu(x)\cdot  \partial_\mu D^\mu(0)\right)
|0 \rangle \right|_{q_\mu\rightarrow 0} =\langle 0| [ -i  Q_D, \partial_\mu D^\mu(0)]|0 \rangle  = \langle 0| -\delta(\partial_\mu D^\mu(0))]|0 \rangle \nonumber \\
&    =4\cdot 
\langle 0|- \frac{\beta^{(\rm NP)}(\alpha)}{4 \alpha} G_{\mu\nu}^2|0 \rangle 
 + \left(3-\gamma_m\right)  \left(1+\gamma_m\right)\cdot  \langle 0|  - \sum^{N_f}_{i=1}   m_f  \bar{\psi}^{i}\psi^{i} |0 \rangle,
  \label{sigmamassformula}  
 \end{eqnarray}
 with the scale dimension $d_{G_{\mu\nu}^2}=4$, $d_{\bar \psi \psi}=3-\gamma_m$. 
Similarly, 
the pole-dominated WT identity for the non-singlet axial-vector current 
$A_\mu^\alpha (\alpha=1,2,3)$ for each doublet  $\psi^i (i=1,2)$  gives the Gell-Mann-Oakes-Renner (GMOR) relation: 
\begin{eqnarray}
F_\pi^2M_\pi^2\cdot  \delta^{\alpha\beta} &=& \langle 0|[ -i Q_5^\alpha,  \partial^\mu A_\mu^\beta(0)]|0 \rangle = \langle  0| - \sum^{2}_{i=1}  m_f   \bar \psi^i  \psi^i |0\rangle \cdot  \delta^{\alpha\beta},
\label{pionmass}
\\ {i.e.} \quad 
\langle 0|  - \sum^{N_f}_{i=1}   m_f  \bar{\psi}^{i}\psi^{i} |0 \rangle&=&\frac{N_f}{2} F_\pi^2 M_\pi^2.
\label{pionmass2}
\end{eqnarray}
 Eq.(\ref{sigmamassformula}) and Eq.(\ref{pionmass}) (usually derived by the soft pion theorem in the broken phase) are simply based on the pole dominance, and hence valid in both the broken and the conformal phases. 
Then we have
~\cite{Matsuzaki:2013eva} 
\begin{eqnarray}
  M_\sigma^2 &=& m_\sigma^2 
          +(3-\gamma_m)(1+\gamma_m) 
          \frac{\frac{N_f}{2} F_\pi^2  M_\pi^2}{F_\sigma^2}, 
          \quad m_\sigma^2\equiv     
        \frac{1}{F_\sigma^2} \langle 0\big| - \frac{\beta^{(\rm NP)}(\alpha)}{\alpha} G_{\mu\nu}^2 \big|0\rangle,  
         \label{eq:msigma-summary}
\end{eqnarray}
{\it independently of the phases}.  
 Any effective theory should reproduce Eq.(\ref{eq:msigma-summary}) for the $\sigma$ mass $M_\sigma^2$.
 
As such  
in the broken phase we use the dilaton ChPT (dChPT) Lagrangian~\cite{Matsuzaki:2013eva} corresponding 
to  Eq.(\ref{thetamumu}):
\begin{eqnarray} 
{\cal L}
&=& {\cal L}
_{\rm inv} + {\cal L}_
{\rm hard}
 + {\cal L}_
{\rm soft}  
\,,
\label{L:p2} \\
 {\cal L}
 _{\rm inv}
&=& \frac{F_\sigma^2}{2} (\partial_\mu \chi)^2 
+ \frac{F_\pi^2}{4} \chi^2 {\rm tr}[\partial_\mu U^\dag \partial^\mu U] ,
\label{inv:part}\\
{\cal L}_{\rm hard}
&=&
-\frac{F_\sigma^2}{4} m_\sigma^2
\chi^4  \left(
 \ln \frac{\chi}{S} - \frac{1}{4}\right), 
 \label{hard:part}\\
 {\cal L}_{\rm soft} &=&{\cal L}^{(1)}
_{\rm soft}+  {\cal L}^{(2)}_{\rm soft}, 
 \nonumber \\
&&{\cal L}^{(1)}_ {\rm soft} =\frac{F_\pi^2}{4} \left( \frac{\chi}{S} \right)^{3-\gamma_m} \cdot S^4 {\rm tr}[ {\cal M}^\dag U + U^\dag {\cal M}] ,
\nonumber \\ 
&&  {\cal L}^{(2)}_{\rm soft} =- \frac{(3-\gamma_m) F_\pi^2}{8}  \chi^4 \cdot {\rm tr}{\cal M}  
\,, 
\label{soft:part}
\end{eqnarray} 
where $U=e^{2i \pi/F_\pi}$, $\chi=e^{\sigma/F_\sigma}$,
${\cal M}$ and $S$ are spurion fields introduced so as to incorporate explicit breaking effects of 
the chiral and scale symmetry, respectively. 
Under the chiral $SU(N_f)_L \times SU(N_f)_R$ symmetry, these building blocks transform as 
$   U \to g_L \cdot U \cdot g_R^\dag $,  
$  {\cal M} \to g_L \cdot {\cal M} \cdot g_R^\dag $, 
$  \chi \to \chi $ and 
$  S \to S $ with $g_{L,R} \in SU(N_f)_{L,R}$, 
while under the scale symmetry they are infinitesimally transformed as  
$  \delta U(x) = x_\nu \partial^\nu U(x) $,  
$  \delta {\cal M}(x) = x_\nu \partial^\nu {\cal M}(x) $,     
$  \delta \chi(x) =  (1 + x_\nu \partial^\nu) \chi(x)  $ and 
$  \delta S = (1 + x_\nu \partial^\nu) S(x) $, with
the vacuum expectation values of 
the spurion fields ${\cal M}$ and $S$, $\langle {\cal M} \rangle= M_\pi^2 \times {\bf 1}_{N_f \times N_f}$ and $\langle S \rangle=1$.

This effective Lagrangian is the same as that of Ref.~\cite{Leung:1989hw} except for  
Eq.(\ref{hard:part}), 
\begin{eqnarray}
  {\cal L}_{\rm hard}&=& -\frac{1}{16} F_\sigma^2}m_\sigma^2 -{\frac{1}{2} m_\sigma^2 \sigma^2
 +\cdots, 
\label{anomaly}
\end{eqnarray}
which is absent in Ref.~\cite{Leung:1989hw}  and gives the $\sigma$ mass in the chiral limit due to the trace anomaly:
  \begin{eqnarray}
  F_\sigma m_\sigma^2&=&<0| \theta_\mu^\mu|_{m_f=0}|\sigma>=<0|\frac{\beta^{(\rm NP)}(\alpha)}{4\alpha} G_{\mu\nu}^2|\sigma>
  =<0|-\delta {\cal L}_{\rm hard} |\sigma>=<0| F_\sigma^2 m_\sigma^2 
\chi^4 \ln \chi|\sigma>  \nonumber\\ 
  &=& 
\frac{4}{F_\sigma} <0| -\theta_\mu^\mu|_{m_f=0}|0>= \frac{1}{F_\sigma} <0|-\frac{\beta^{(\rm NP)}(\alpha)}{\alpha} G_{\mu\nu}^2|0> 
=\frac{16}{F_\sigma} <0| {\cal L}_{\rm hard} |0> , 
\label{anomaly2}
  \end{eqnarray}
to be compared with Eq.(\ref{eq:msigma-summary}).
On the other hand, ${\cal L}_{\rm soft}$ has two terms: ${\cal L}_{\rm soft}^{(1)}$ corresponds to the fermion mass term~\cite{Leung:1989hw}:
\begin{eqnarray}
\sum^{N_f}_{i=1} m_f \bar \psi_i \psi_i =
- {\cal L}_{\rm soft}^{(1)}&=& - \frac{F_\pi^2}{4} \left( \chi \right)^{3-\gamma_m} \cdot  {\rm tr}[ {\cal M}^\dag U + U^\dag {\cal M}] \nonumber\\
 &=&\left[1+(3-\gamma_m) \frac{\sigma}{F_\sigma} + \frac{1}{2} (3-\gamma_m)^2 \frac{\sigma^2}{F_\sigma^2}\right]
 \left(- \frac{N_f}{2} F_\pi^2  M_\pi^2 +  \frac{M_\pi^2}{2}\pi^a \pi^a\right) +\cdots\,
 \label{fermionmassterm}
   \end{eqnarray}
which correctly reproduces the $\pi$ mass term as in the standard ChPT, $\frac{M_\pi^2}{2}\pi^a \pi^a$, and  the GMOR relation Eq.(\ref{pionmass}),
   but would imply that $\sigma$ is a tachyon  to destabilize the vacuum (in the case $m_\sigma^2=0$):
 $M_\sigma^2~= - (3-\gamma_m)^2 \frac{N_f}{2} F_\pi^2 M_\pi^2 /F_\sigma^2 <0$. Then ${\cal L}^{(2)}_{\rm soft}$ was introduced in Ref.~\cite{Leung:1989hw} to avoid the $\sigma$ to be a tachyon:
\begin{eqnarray}
- {\cal L}^{(2)}_{\rm soft}= \frac{(3-\gamma_m) F_\pi^2}{8}  \chi^4 \cdot {\rm tr}{\cal M}  = \frac{3-\gamma_m}{4} \left(1+4\frac{\sigma}{F_{\sigma} }
+8 \frac{\sigma^2}{F_\sigma^2}\right) \frac{N_f}{2} F_\pi^2 M_\pi^2 +\cdots,
\label{sigmamasscontribution}
\end{eqnarray}
which is essential for the correct $\sigma$ mass term 
$- M_\sigma^2 \sigma^2/2$ (in addition to $-m_\sigma^2\sigma^2/2$ in Eq. (\ref{anomaly})) 
given as a combination of the two terms of ${\cal L}_{\rm soft}$:
\begin{eqnarray}
M_\sigma^2 = m_\sigma^2 +\left[-(3-\gamma_m)^2 +4(3-\gamma_m)\right] \frac{\frac{N_f}{2} F_\pi^2 M_\pi^2}{F_\sigma^2}= m_\sigma^2 +
 (3-\gamma_m) (1+\gamma_m)\frac{ \frac{N_f}{2} F_\pi^2 M_\pi^2}{F_\sigma^2},
 \label{sigmamassterm}
 \end{eqnarray}
thus correctly reproduces  $\sigma$ mass formula derived by the WT identity Eq.(\ref{eq:msigma-summary}).

The same mass formula is also obtained through the trace of the 
energy-momentum tensor $<0| \theta_\mu^\mu|\sigma>=M_\sigma^2 F_\sigma$:
 \begin{eqnarray}
( M_\sigma^2 -m_\sigma^2) F_\sigma  
&=&<0|   (1+\gamma_m) \sum^{N_f}_{i=1} m_f \bar \psi_i \psi_i |\sigma>
=(1+\gamma_m)<0|  - \delta ( \sum^{N_f}_{i=1} m_f \bar \psi_i \psi_i )|0>  \nonumber\\ 
&=&
<0|- \delta {\cal L}_{\rm soft}|\sigma>
= (1+\gamma_m)\cdot (3-\gamma_m)
\frac{ \frac{N_f}{2} F_\pi M_\pi^2}{F_\sigma^2},
\label{Msigmamass} 
\end{eqnarray}
where $ <0|  - \delta ( \sum^{N_f}_{i=1} m_f \bar \psi_i \psi_i )|0> = (3-\gamma_m) <0| - \sum^{N_f}_{i=1} m_f \bar \psi_i \psi_i )|0> $ with Eq.(\ref{pionmass2}), while 
\begin{eqnarray}
-\delta {\cal L}_{\rm soft}&=&-(\delta {\cal L}^{(1)}_{\rm soft} + \delta {\cal L}^{(2)}_{\rm soft} )=
(3-\gamma_m)\chi^{3-\gamma_m} \left[-  \frac{N_f}{2} F_\pi^2  M_\pi^2  +\frac{M_\pi^2}{2}\pi^a \pi^a\right]
+  (3-\gamma_m) \chi^4
  \frac{N_f}{2} F_\pi^2  M_\pi^2 
 + \cdots\nonumber\\
 &=&   (3-\gamma_m) \left[- (3-\gamma_m)+4\right] \frac{N_f}{2} F_\pi^2  M_\pi^2\frac{\sigma}{F_\sigma} - (3-\gamma_m)
 \frac{M_\pi^2}{2}\pi^a \pi^a
 + \cdots , 
\label{deltaLsoft} 
 \end{eqnarray}
 both giving the same result. 
This is compared with 
$\sum^{N_f}_{i=1} m_f \bar \psi_i \psi_i $ in Eq.(\ref{fermionmassterm}) having no contribution of ${\cal L}^{(2)}_{\rm soft}$:
\begin{eqnarray}
<0|2 \cdot \sum^{N_f}_{i=1} m_f \bar \psi_i \psi_i|\sigma>  &=&2 \cdot < 0|-\delta (\sum^{N_f}_{i=1} m_f \bar \psi_i \psi_i)|0>\frac{1}{F_\sigma}
= 2 \cdot <0|\delta {\cal L}_{\rm soft}^{(1)}|0>\frac{1}{F_\sigma} \nonumber\\
&=& 
2 \cdot (3-\gamma_m)  
\frac{ \frac{N_f}{2} F_\pi^2 M_\pi^2}{F_\sigma}
= \frac{2}{1+\gamma_m}\cdot  (M_\sigma^2-m_\sigma^2)F_\sigma. 
\label{2fermionmassterm}
\end{eqnarray}
Eqs.(\ref{Msigmamass}) and (\ref{2fermionmassterm}) are crucial to compare later the $\sigma$ pole contribution to the $<\pi| (1+\gamma_m) \cdot  \sum^{N_f}_{i=1} m_f \bar \psi_i \psi_i |\pi>$ and $  <\pi| 2\cdot \sum^{N_f}_{i=1} m_f \bar \psi_i \psi_i |\pi>$, respectively.

The result Eq.(\ref{sigmamassterm}) (and (\ref{Msigmamass}))
coincides with that of Ref.~\cite{Leung:1989hw} for $m_\sigma^2=0$.~\footnote{Thus Ref.~\cite{Leung:1989hw} 
implicitly assume $\frac{\beta(\alpha)}{4\alpha} G_{\mu\nu}^2=0$, or $m_\sigma^2=0$, in the broken phase, which is in contradiction to their own calculation by the ladder SD equation which shows no massless dilaton in the chiral limit, see also Ref.\cite{Matsuzaki:2015sya}.} Zwicky also assumes $m_\sigma^2=0$. He evaluated  $\left<\pi(p_2)|(1+\gamma_m) \cdot   \sum^{N_f}_{i=1} m_f \bar \psi_i \psi_i|\pi(p_1)\right>$ through  the form factor argument on $\left<\pi(p_2)|\theta_\mu^\mu |\pi(p_1)\right>
=\left<\pi(p_2)|\frac{\beta(\alpha)}{4\alpha} G_{\mu\nu}^2+(1+\gamma_m)    \sum^{N_f}_{i=1} m_f \bar \psi_i \psi_i |\pi(p_1) \right>$,
which is known to give $2 M_\pi^2$ at $q^2=(p_1-p_2)^2 \rightarrow 0$. Then he needed the assumption of the existence of the IR fixed point (in the broken phase with $M^2_\pi\ne0$) in order to drop out the contribution of $ \left<\pi(p_2)|\frac{\beta(\alpha)}{4\alpha} G_{\mu\nu}^2 |\pi(p_1)\right>(\propto m_\sigma^2)$ to conclude $\left<\pi(p_2)|(1+\gamma_m) \cdot   \sum^{N_f}_{i=1} m_f \bar \psi_i \psi_i|\pi(p_1)\right>=2 M_\pi^2 $.

However, this term is actually irrelevant to the discussion here to directly compute 
$\left<\pi(p_2)|(1+\gamma_m) \cdot   \sum^{N_f}_{i=1} m_f \bar \psi_i \psi_i|\pi(p_1)\right>$ by dChPT without referring to $\left<\pi(p_2)|\theta_\mu^\mu|\pi(p_1)\right>$, and then compare it with $\left<\pi(p_2)| 2 \cdot   \sum^{N_f}_{i=1} m_f \bar \psi_i \psi_i|\pi(p_1)\right>$ computed on the same footing based on the dChPT. 
Anyway, our result with Eqs.(\ref{Msigmamass}) and (\ref{2fermionmassterm}) obviously shows the same conclusion even including the trace anomaly, $m_\sigma^2\ne 0$.
Hence the following discussion is irrelevant to the Zwicky's assumption that there exists  the IR fixed point,  $\frac{\beta(\alpha)}{4\alpha} G_{\mu\nu}^2=0$
even in the broken phase with the condensate $\left< \bar \psi \psi \right>|_{m_f=0} \ne 0$ (its mass scale is the explicit as well as spontaneous breaking of the scale symmetry) and $M_\pi^2\ne 0$ (explicit breaking of the scale symmetry as well as chiral symmetry).  Anyway such an assumption itself  has been shown to be in contradiction with the explicit calculation in the ladder SD equation (see footnote \ref{fixedpoint})~\cite{Matsuzaki:2015sya}.

\section{Evaluation of matrix element between pion states on the mass shell}

Before evaluation by the dChPT Lagrangian, we first see the generic argument for $\left<\pi(p_2)|\theta^{\mu\nu}(x^\mu=0)|\pi(p_1)\right>$ based on the form factor:
\begin{eqnarray}
\left<\pi(p_2)|\theta^{\mu\nu}|\pi(p_1)\right>
&=&2P^\mu P^\nu F(q^2) + (g^{\mu\nu} q^2 - q^\mu q^\nu) G(q^2), \nonumber\\
P^\mu=(p_1^\mu+p_2^\mu)/2, \,\, q^\mu&=& p_2^\mu-p_1^\mu,\,\,F(0)=1, \,\,G(q^2)|_{M_\sigma^2\ne 0} \,\, {\rm regular\, at} \,\, 
q^2 \rightarrow 0 \\
\left<\pi(p_2)|\theta^\mu_\mu |\pi(p_1) \right>
&=& 2 M_\pi^2 F(q^2) + q^2 \left[ 3 G(q^2) -F(q^2)/2\right]\,\nonumber\\
&\rightarrow& 2 M_\pi^2 \quad {\rm at} \quad q^2 \rightarrow 0.
\label{FFargument}
\end{eqnarray}
It should be noted that in this formula the $\sigma$ pole contribution is invisible at $q^2\rightarrow 0$ and the result  is valid independently of the phases, either the broken phase or the conformal phase.

Now we evaluate the same quantity through the dChPT Lagrangian Eq.(\ref{L:p2})  for the broken phase~\cite{Leung:1989hw}:
\begin{eqnarray}
\left<\pi(p_2)|\theta_\mu^\mu|\pi(p_1)\right>
&=& 4 M_\pi^2 - 2 p_1\cdot p_2 + \left<0|\theta_\mu^\mu|\sigma(q)\right>\frac{1}{M_\sigma^2-q^2}  G_{\sigma \pi\pi} (q^2, M_\pi^2,M_\pi^2)\, \nonumber\\
&=&2 M_\pi^2 + q^2 + \frac{q^2}{M_\sigma^2-q^2}\left[(1-\gamma_m) M_\pi^2 + q^2\right], 
\label{thetamumuvalue}
\end{eqnarray}
where $\left<0|\theta_\mu^\mu|\sigma(q)\right>=F_\sigma q^2$ and 
\begin{eqnarray}
F_\sigma G_{\sigma \pi\pi} (q^2, M_\pi^2,M_\pi^2)&=& (3-\gamma_m) M_\pi^2 - 2 p_1\cdot p_2 =(1-\gamma_m) M_\pi^2 + q^2, 
\label{sigmapipi}
\end{eqnarray}
with the $\sigma-\pi-\pi$ vertex $G_{\sigma \pi\pi} (q^2, M_\pi^2,M_\pi^2)$ given by $F_\sigma G_{\sigma \pi\pi} (q^2, M_\pi^2,M_\pi^2)=(1-\gamma_m) M_\pi^2 + q^2$  as  a sum of $(3-\gamma_m) M_\pi^2$ from explicit breaking term Eq.(\ref{soft:part}) and $- 2 p_1\cdot p_2=q^2- 2M_\pi^2$ from the pion kinetic term in Eq.(\ref{inv:part}). (Eq.(\ref{sigmapipi}) was also obtained in Ref.\cite{Ellis:1970yd} in a different context.)
Eq.(\ref{thetamumuvalue}) is consistent with the form factor argument Eq.(\ref{FFargument}):
\begin{eqnarray}
\left<\pi(p_2)|\theta_\mu^\mu|\pi(p_1)\right>
 \rightarrow 2 M_\pi^2\,\quad  {\rm at}\,\, q^2 \rightarrow 0,
\end{eqnarray}
again with the $\sigma$ pole contribution being invisible at $q^2 \rightarrow 0$. Note that this  implies that the trace anomaly term giving $m_\sigma^2 \ne0$ does not contribute  to $\left<\pi(p_2)|\theta_\mu^\mu|\pi(p_1)\right>$  at $q^2 \rightarrow 0$, even without assumption of IR fixed point.

On the other hand, we have:
\begin{eqnarray}
&&\left<\pi(p_2)|(1+\gamma_m) \cdot   \sum^{N_f}_{i=1} m_f \bar \psi_i \psi_i|\pi(p_1)\right>
=\left<\pi(p_2)|  
 -\delta {\cal L}_{\rm inv} -\delta
 {\cal L}_{\rm soft} |\pi(p_1)\right>
   \nonumber\\
 &&= [4-(3-\gamma_m) ] M_\pi^2 +
<0|-\delta {\cal L}_{\rm soft}|\sigma> \frac{1}{M_\sigma^2-q^2} G_{\sigma\pi\pi} 
=(1+\gamma_m) M_\pi^2 +\frac{M_\sigma^2- m_\sigma^2}{M_\sigma^2-q^2} \left[(1-\gamma_m) M_\pi^2 + q^2 \right] \nonumber \\
&&=
\left[2 M_\pi^2 +q^2 +\frac{q^2}{M_\sigma^2-q^2} \left[(1-\gamma_m) M_\pi^2 + q^2 \right] \right]
-\frac{m_\sigma^2}{M_\sigma^2-q^2} \left[(1-\gamma_m) M_\pi^2 + q^2 \right], 
\label{massbreaking}
\end{eqnarray}
where use has been made of  Eqs.(\ref{Msigmamass},\ref{deltaLsoft}) and Eq.(\ref{sigmapipi}). Note that the $\sigma$ pole term of Eq.(\ref{massbreaking})  is from the pole of $\sigma$ in the scalar density $\bar \psi \psi$ coupled to 2 $\pi$'s,
with the $\sigma-\pi-\pi$ coupling $F_\sigma G_{\sigma \pi\pi} (q^2, M_\pi^2,M_\pi^2)$ in Eq.(\ref{sigmapipi}).  
Eq.(\ref{massbreaking}) is identical to Eq.(\ref{thetamumuvalue}), with the last term  being precisely  the same as the $\sigma$ pole  contribution to the trace anomaly, Eq.(\ref{anomaly2}):
\begin{eqnarray}
 <\pi(p_2)|\frac{\beta^{(\rm NP)}(\alpha)}{4\alpha} G_{\mu\nu}^2|\pi(p_1)>=  \frac{<0|\frac{\beta^{(\rm NP)}(\alpha)}{4\alpha} G_{\mu\nu}^2|\sigma>}{M_\sigma^2-q^2} G_{\sigma \pi\pi} (q^2, M_\pi^2,M_\pi^2)
=\frac{m_\sigma^2}{M_\sigma^2-q^2} \left[(1-\gamma_m) M_\pi^2 + q^2 \right],
\end{eqnarray}
to be cancelled each other for $<\pi(p_2)|\theta_\mu^\mu|\pi(p_1)>$ in Eq.(\ref{thetamumuvalue}).
At $q^2\rightarrow 0$ we have
\begin{eqnarray}
<\pi(p_2)|(1+\gamma_m) \cdot   \sum^{N_f}_{i=1} m_f \bar \psi_i \psi_i|\pi(p_1)>
&=& 2 M_\pi^2  -\frac{m_\sigma^2}{M_\sigma^2} (1-\gamma_m) M_\pi^2\quad
{\rm at} \,\, q^2\rightarrow 0\,.
\label{polecontribution}
 \end{eqnarray}

Now to the matrix element $<\pi(p_2)|2 \cdot   \sum^{N_f}_{i=1} m_f \bar \psi_i \psi_i|\pi(p_1)>$. 
From  Eq.(\ref{2fermionmassterm}) we have:~\footnote{
In contrast, if we used the conventional non-scale-invariant ChPT Lagrangian with Eq.(\ref{fermionmassterm}) replaced by that ignoring the $\sigma$ terms, then the RHS would be just 
$2 M_\pi^2$, which is also obtained by the double soft pion theorem as claimed of Zwicky. See also the later discussion.
}
\begin{eqnarray}
\left<\pi(p_2)| 2 \cdot   \sum^{N_f}_{i=1} m_f \bar \psi_i \psi_i|\pi(p_1)\right>
&=& 2\cdot <\pi(p_2)|-{\cal L}_{\rm soft}^{(1)}   |\pi(p_1)>
\nonumber\\
&=& 2M_\pi^2+ 2  <0|\delta {\cal L}_{\rm soft}^{(1)} |0>  
 \frac{1}{M_\sigma^2-q^2} G_{\sigma\pi\pi} \nonumber\\
&=&2 M_\pi^2+\frac{2}{1+\gamma_m} \frac{M_\sigma^2- m_\sigma^2}{M_\sigma^2-q^2} \left[(1-\gamma_m) M_\pi^2 + q^2 \right] \nonumber\\
&=&\frac{2}{1+\gamma_m}\cdot 2 M_\pi^2  - \frac{2}{1+\gamma_m}\frac{m_\sigma^2}{M_\sigma^2} (1-\gamma_m) M_\pi^2\quad {\rm at} \,\, q^2\rightarrow 0,
\label{polecontribution2}
\end{eqnarray}
where Eq.(\ref{2fermionmassterm}) was used.

From Eqs.(\ref{polecontribution}) and (\ref{polecontribution2}) we conclude: 
\begin{eqnarray}
<\pi(p_2)|2 \cdot   \sum^{N_f}_{i=1} m_f \bar \psi_i \psi_i|\pi(p_1)>
 &=& \frac{2}{1+\gamma_m} <\pi(p_2)| (1+\gamma_m) \cdot   \sum^{N_f}_{i=1} m_f \bar \psi_i \psi_i|\pi(p_1)>,
\nonumber \\
&\ne&  <\pi(p_2)| (1+\gamma_m) \cdot   \sum^{N_f}_{i=1} m_f \bar \psi_i \psi_i|\pi(p_1)>, 
\end{eqnarray}
thus no constraint on $\gamma_m$, in contradiction to Zwicky's claim, even including the trace anomaly $m_\sigma^2 \ne 0$, hence  independently of the IR fixed point argument.

The crucial point of the results  is the contribution of the pole of $\sigma$, $(1-\gamma_m)M_\pi^2$ and $2\frac{1-\gamma_m}{1+\gamma_m} M_\pi^2$,  without which we would erroneously conclude 
  $\left<\pi(p_2)|(1+\gamma_m) \cdot   \sum^{N_f}_{i=1} m_f \bar \psi_i \psi_i |\pi(p_1)\right> 
  =(1+\gamma_m) M_\pi^2$ (as emphasized in Ref.\cite{Leung:1989hw}) and $\left<\pi(p_2)| 2 \cdot   \sum^{N_f}_{i=1} m_f \bar \psi_i \psi_i|\pi(p_1)\right> 
  = 2M_\pi^2$, compared with the correct ones, $2 M_\pi^2$ and $\frac{2}{1+\gamma_m} \cdot 2M_\pi^2 $,  respectively (up to trace anomaly $m_\sigma^2$ term).
This implies that Zwicky's  argument corresponds to the inclusion of the $\sigma$ pole for the former, while neglect for the latter.
\\

In fact, Zwicky's  arguments (assuming $m_\sigma^2=0$) are equivalent to the {\it neglect of the $\sigma$ pole} contribution in Eq.(\ref{polecontribution2}) to arrive at $\left<\pi(p_2)| 2\cdot   \sum^{N_f}_{i=1} m_f \bar \psi_i \psi_i |\pi(p_1)\right>= 2 M_\pi^2$,  
which he in fact showed to be
equivalent to the double use of the soft pion theorem (unjustifiably {\it removing  the $\sigma$ pole contribution})~\footnote{If we use the soft pion theorem 
$\left<\pi(p_2)| 2 \cdot   \sum^{N_f}_{i=1} m_f \bar \psi_i \psi_i|\pi(p_1)\right>|_{p_1^2=M_\pi^2, p_2\rightarrow 0}= \left<0| \left[i Q_5^a, 2 \cdot   \sum^{N_f}_{i=1} m_f \bar \psi_i \psi_i\right] |\pi(p_1)\right>|_{p_1^2=M_\pi^2, p_2\rightarrow 0}/F_\pi$, the resultant expression removes the 
$\sigma$ pole contribution, since  $\sigma$ is a chiral singlet, $[i Q_5^a, \sigma]=0$ (dilaton $\sigma$ is different from the ``sigma'' ($\hat \sigma$) in the linear sigma model which is a chiral partner of ${\hat \pi}^a$, with the correspondence to $\sigma$ as~\cite{Yamawaki:2016qux}:
${\hat \sigma}^2 + ({\hat \pi}^a)^2=(F_\pi \cdot \chi)^2= F_\pi^2\cdot  e^{2\sigma/F_\pi}$).
}. 
On the other hand, for $\left<\pi(p_2)|(1+\gamma_m) \cdot   \sum^{N_f}_{i=1} m_f \bar \psi_i \psi_i |\pi(p_1)\right>$ he equated it with  the generic result Eq.(\ref{thetamumuvalue}) (where $\sigma$ pole is  invisible at $q^2 \rightarrow 0$), although the same result   $\left<\pi(p_2)|(1+\gamma_m) \cdot   \sum^{N_f}_{i=1} m_f \bar \psi_i \psi_i |\pi(p_1)\right>
 = 2 M_\pi^2$ through the direct computation is obtained 
{\it only when including the $\sigma$ pole},  as shown in Eq.(\ref{polecontribution}). Equating the two results dealing $\sigma$ pole differently,  he concluded $2=1+\gamma_m$, i.e.,  $\gamma_m=1$. 
 
Putting differently, we may consistently use the same double soft pion theorem on both of $\left<\pi(p_2)| (1+\gamma_m)\cdot    \sum^{N_f}_{i=1} m_f \bar \psi_i \psi_i |\pi(p_1)\right>$and $\left<\pi(p_2)| 2 \cdot  \sum^{N_f}_{i=1} m_f \bar \psi_i \psi_i |\pi(p_1)\right>$ (though both $\pi$'s are not on the mass shell in contrast to the main stream of the present discussion),  
which implies {\it neglecting $\sigma$ pole for both}.  By this we would get  
\begin{eqnarray}
<\pi(p_2)| (1+\gamma_m) \cdot   \sum^{N_f}_{i=1} m_f \bar \psi_i \psi_i |\pi(p_1)
>|_{p_1,p_2\rightarrow 0} &=&
(1+\gamma_m) \cdot 
<0|  -  \sum^{2}_{i=1} m_f \bar \psi_i \psi_i |0
>/F_\pi^2=(1+\gamma_m) M_\pi^2\nonumber, \\ 
<\pi(p_2)| 2\cdot   \sum^{N_f}_{i=1} m_f \bar \psi_i \psi_i |\pi(p_1)
>|_{p_1,p_2\rightarrow 0}&=&2 \cdot 
<0| -\sum^{2}_{i=1} m_f \bar \psi_i \psi_i |0
>/F_\pi^2 =2 M_\pi^2
\label{doublesoft}
\end{eqnarray}
which coincides with the result neglecting the $\sigma$ pole contributions in Eq.(\ref{massbreaking}) and (\ref{polecontribution2}), where GMOR relation $M_\pi^2= - \left<0| \sum^{2}_{i=1} m_f \bar \psi_i \psi_i|0\right>/F_\pi^2$, Eq.(\ref{pionmass}), was used.
\footnote{
GMOR relation is based on the single use of the soft pion theorem for the {\it axialvector current}  which has no pole of $\sigma$, while the {\it flavor-singlet scalar density} $\bar \psi \psi$ has the same quantum number as $\sigma$ and both $(1+\gamma_m) \bar \psi \psi$ and $2 \bar \psi \psi$ equally have a $\sigma$ pole with the coupling to two $\pi$'s given in  Eq.(\ref{sigmapipi}). Different from GMOR, the double use of the soft pion theorem for $\bar \psi \psi$ ignoring the $\sigma$ pole contribution is not justified.  %= 2 \cdot M_\pi^2$.  %,and hence $1+\gamma_m=2$.
} 
Thus again
$<\pi(p_2)| (1+\gamma_m)  \cdot  \sum^{N_f}_{i=1} m_f \bar \psi_i \psi_i |\pi(p_1)>
\ne  <\pi(p_2)| 2\cdot   \sum^{N_f}_{i=1} m_f \bar \psi_i \psi_i |\pi(p_1)>$,
namely, no constraint on the value of $\gamma_m$ (or $\gamma_*$) in contradiction to Zwicky's argument claiming both sides equally to be $2M_\pi^2$.

Of course, the inequality  is trivially true, with the same matrix element 
$\left<\pi(p_2)|  \sum^{N_f}_{i=1} m_f \bar \psi_i \psi_i |\pi(p_1)\right>|_{p_1,p_2\rightarrow 0}$ evaluated by the same method, is simply multiplied by the different numerical factor $1+\gamma_m$ vs $2$. 
However, the message of this trivial game is as follows:
The double use of the soft pion theorem for the scalar density (coupled to $\sigma$) simply misses the (massive) $\sigma$ pole contribution  $(1-\gamma_m) M_\pi^2$ for $\left<\pi(p_2)| (1+\gamma_m) \cdot  \sum^{N_f}_{i=1} m_f \bar \psi_i \psi_i |\pi(p_1)\right>$, inclusion of which gives the correct results, $2 M_\pi^2$, consistent with the form factor argument as shown in Eq.(\ref{polecontribution}), while including the $\sigma$ pole also in  $\left<\pi(p_2)| 2\cdot  \sum^{N_f}_{i=1} m_f \bar \psi_i \psi_i |\pi(p_1)\right>$ would no longer keep $2 M_\pi^2$, 
actually $2/(1+\gamma_m) \cdot 2M_\pi^2$, thus again arriving at  inequality, when $\sigma$ pole 
is included in both consistently, i.e., the equality $1+\gamma_m=2$ is lost anyway. 

More strikingly, double use of soft pion theorem also implies $\left<\pi(p_2)| \theta_\mu^\mu |\pi(p_1)\right>|_{p_1,p_2\rightarrow 0}=\left<\pi(p_2)| (1+\gamma_m)    \sum^{N_f}_{i=1} m_f \bar \psi_i \psi_i |\pi(p_1)\right>|_{p_1,p_2\rightarrow 0}$, since the trace anomaly term $ \beta(\alpha)/(4\alpha) G_{\mu\nu}^2$ is a chiral singlet, $\left[i Q_5^a,\beta(\alpha)/(4\alpha) G_{\mu\nu}^2\right]=0$,  and the soft pion theorem makes its contribution zero, {\it independently of the Zwicky's assumption of the IR fixed point}.
Hence we would get $\left<\pi(p_2)| \theta_\mu^\mu |\pi(p_1)\right>|_{p_1,p_2\rightarrow 0}= (1+\gamma_m) M_\pi^2  \ne 2 M_\pi^2$, in contradiction with the form factor argument which  the Zwicky's arguments are crucially based on.
\\

 So far the dChPT result. We now comment on  the Zwicky's argument based on the Feynman-Hellmann theorem, Eq(2.20) in Ref.\cite{Zwicky:2023bzk}:
\begin{eqnarray}
\left<\pi(p_2)| 2 \cdot   \sum^{N_f}_{i=1} m_f \bar \psi_i \psi_i|\pi(p_1)\right>
&=&2 \frac{\partial}{\partial \ln m_f}  \left<\pi(p_2)| {\cal H}|\pi(p_1) \right> 
=\frac{\partial}{\partial \ln m_f} (2 E_\pi\cdot E_\pi) 
\nonumber\\
&=&
\frac{\partial}{\partial \ln m_f} 2 M_\pi^2 = 2 M_\pi^2, 
\label{hamiltonian}
\end{eqnarray}
up to order of $m_f^2$.  The last equation depends crucially on {\it  his assumption of the combined use of  $M_\pi^2 \sim m_f$} which is  characteristic to the pion as a pseudo NG boson in the broken phase.
However, if we used the same theorem to $  \left<\pi(p_2)| (1+\gamma_m) \cdot   \sum^{N_f}_{i=1} m_f \bar \psi_i \psi_i|\pi(p_1)\right>= (1 +\gamma_m) \frac{\partial}{\partial \ln m_f} \left<\pi(p_2)| {\cal H}|\pi(p_1) \right>$ 
with the same assumption $M_\pi^2 \sim m_f$,
then we would get  $  \left<\pi(p_2)| (1+\gamma_m) \cdot   \sum^{N_f}_{i=1} m_f \bar \psi_i \psi_i|\pi(p_1)\right> = (1+\gamma_m) M_\pi^2\ne 2 M_\pi^2 $ in contradiction with the generic result  in Eq.(\ref{FFargument}).
The theorem is insensitive to the spontaneous symmetry breaking, giving the same form in $M_\pi^2$ before taking derivative $\frac{\partial}{\partial \ln m_f} $ for both the broken phase and conformal phase.

 Actually, if we apply the same  theorem to the conformal phase  where dChPT is invalid and without $\sigma$ pole contribution, 
 we may use the hyperscaling $M_\pi \sim m_f^{1/(1+\gamma_m)}$ 
 to get
 \begin{eqnarray}
 \left<\pi(p_2)| 2 \cdot   \sum^{N_f}_{i=1} m_f \bar \psi_i \psi_i|\pi(p_1)\right> 
 &=&2\frac{\partial}{\partial \ln m_f}  \left<\pi(p_2)| {\cal H}|\pi(p_1) \right>
 = \frac{\partial}{\partial \ln m_f} 2 E_\pi^2
 = \frac{2}{1+\gamma_m} \cdot 2 M_\pi^2,\nonumber\\
 \left<\pi(p_2)| (1+\gamma_m) \cdot   \sum^{N_f}_{i=1} m_f \bar \psi_i \psi_i|\pi(p_1)\right>
 &=& 
 (1+\gamma_m) \frac{\partial}{\partial \ln m_f}  \left<\pi(p_2)| {\cal H}|\pi(p_1) \right>
  = 2 M_\pi^2,
   \label{FHtheorem2}
  \end{eqnarray}
with the latter now being consistent with the generic phase-independent result in Eq.(\ref{FFargument}) as it should be.
Eq.(\ref{FHtheorem2}) is   the same result as in the broken phase through  the dChPT,  up to the trace anomaly term $m_\sigma^2 \ne 0$ (which is the pole term).
It is curious that the combined use of the Feynman-Hellmann theorem and $M_\pi^2\sim m_f$ coincides with the wrong result of the double-soft pion theorem Eq.(\ref{doublesoft}) ignoring the $\sigma$ pole contribution, 
while combined use of the hyperscaling (followed by the simple Coulombic bound state) even for the pion in the broken phase gives the correct result phase-independently.

   \section{Additional Comments}
   1) $\sigma-\pi-\pi$ vertex in Eq.(\ref{sigmapipi})   
   
  One might be concerned about the $\sigma-\pi-\pi$ vertex in Eq.(\ref{sigmapipi}). It is different from the well-known low energy theorem of the scale symmetry~\cite{Yamawaki:2016qux}
,  
  \begin{eqnarray}
  F_\sigma G_{\sigma\pi\pi}(q^2, M_\pi^2, M_\pi^2)=2 M_\pi^2, \quad q^2 \rightarrow 0,
  \label{LETcoupling}
  \end{eqnarray}
 which is also obtained by the dispersion representation, $\left<\pi(p_2)|\theta_\mu^\mu|\pi(p_1)\right>=M_\sigma^2/(M_\sigma^2-q^2)\cdot F_\sigma G_{\sigma\pi\pi} (q^2,M_\pi^2,M_\pi^2)$, compared  with the form factor argument Eq.(\ref{FFargument}). 
Both are valid for the $\sigma$ as a  pseudo-dilaton but  {\it $\pi$  as a non NG boson} (massive matter field) like $\rho$ meson.  
 
  On the other hand, Eq.(\ref{sigmapipi})  
  is the result for the case  of {\it both $\sigma$ and $\pi$ being pseudo-NG bosons}, since it is  a sum of $(3-\gamma_m) M_\pi^2$ from explicit breaking term Eq.(\ref{soft:part}) and $- 2 p_1\cdot p_2=q^2- 2M_\pi^2$ from the pion kinetic term Eq.(\ref{inv:part}), both characteristic to the spontaneously broken scale and chiral symmetries for $\sigma$ and $\pi$.  
At $q^2=0$ it reads $F_\sigma G_{\sigma\pi\pi}(0,M_\pi^2, M_\pi^2)=(1-\gamma_m)M_\pi^2$, obviously different from the low energy theorem of the scale symmetry. 
\\

2) $f_0(500)$ meson for $N_f=2$ as a massive dilaton

John Ellis \cite{Ellis:1970yd} obtained the same result as Eq.(\ref{sigmapipi}): $F_\sigma G_{\sigma\pi\pi}(q^2,M_\pi^2,M_\pi^2)= - \lambda M_\pi^2 -2 p_1\cdot p_2 = (1-\gamma_m)M_\pi^2 + q^2$, with $\lambda= -(3-\gamma_m)$.
He instead focused on the  on-shell $\sigma$ coupling $F_\sigma G_{\sigma\pi\pi}(M_\sigma^2,M_\pi^2,M_\pi^2)=(1-\gamma_m) M_\pi^2 + M_\sigma^2$, however with $M_\sigma^2$ free parameter. 

It is compared with our case where $M_\sigma$ is not a free parameter but is constrained as~\cite{Matsuzaki:2013eva};  
\begin{eqnarray}
M_\sigma^2
=m_\sigma^2+ (3-\gamma_m)(1+\gamma_m) (N_f/2)(F_\pi^2/F_\sigma^2) \cdot M_\pi^2,
\end{eqnarray}
which is derived not only through the dChPT Lagrangian {Eq.(\ref{L:p2}) valid in the broken phase, Eq.(\ref{sigmamassterm}),  but also more generally through the WT identity, Eq.(\ref{eq:msigma-summary}),  and hence valid both for broken phase and conformal phases.
  Were it not for $m_\sigma^2=-\left<0|\beta(\alpha)/(\alpha) G_{\mu\nu}^2|0 \right>/F_\sigma^2$ as in the Zwicky's case, we would have $M_\sigma^2= {\cal O} (M_\pi^2)$ for $(3-\gamma_m)(1+\gamma_m) \cdot (N_f/2)(F_\pi^2/F_\sigma^2) ={\cal O} (1)$, and hence $F_\sigma G_{\sigma\pi\pi}(M_\sigma^2,M_\pi^2, M_\pi^2)={\cal O} ( M_\pi^2)$, roughly the same as the low energy theorem for $\sigma$: $F_\sigma G_{\sigma\pi\pi}(0,M_\pi^2, M_\pi^2)= 2 M_\pi^2$. In fact $N_f=8$ LatKMI data~\cite{LatKMI:2016xxi} read $M^2_\sigma \simeq M_\pi^2\gg m_\sigma^2$ and hence $F_\sigma G_{\sigma\pi\pi}(M_\sigma^2,M_\pi^2, M_\pi^2) \simeq  M_\sigma^2\simeq M_\pi^2$.
 
 On the other hand, the real $N_f=2$ QCD in the deep broken phase near the chiral limit, the $\sigma$ mass should be mainly due to the trace anomaly $ m_\sigma^2=- \left<0|\beta(\alpha)/(\alpha) G_{\mu\nu}^2|0 \right> /F_\sigma^2  \gg M_\pi^2$, such that $M_\sigma^2 \simeq m_\sigma^2\gg M_\pi^2$ suggesting identification $\sigma$ as $f_0(500)$. Then, thanks to the trace anomaly dominance  in the mass formula above, the formula Eq.(\ref{sigmapipi}) definitely predicts  $\sigma-\pi-\pi$ coupling for the $\sigma$ on the mass shell $q^2=M_\sigma^2 \, (\gg (1-\gamma_m) M_\pi^2)$: 
  \begin{eqnarray}
 G_{\sigma\pi\pi}(M_\sigma^2,M_\pi^2, M_\pi^2) \simeq M_\sigma^2/F_\sigma \simeq M_\sigma^2/F_\pi \gg 2 M_\pi^2/F_\pi,
 \end{eqnarray}
 with $F_\sigma \simeq F_\pi$.  If it is the case,  the width of $f_0(500)$  will be enhanced by $[M_\sigma^2/(2 M_\pi^2)]^2 \sim 50$ times large than the low theorem value in Eq.(\ref{LETcoupling}), in rough agreement with the reality and  $f_0(500)$ may be regarded as a pseudo NG boson, pseudo dilaton (though very massive far from the invariant limit).
 The crucial point is that in addition to the {\it dominance of the non-perturbative trace anomaly for $M_\sigma^2$}, the formula Eq.(\ref{sigmapipi}) for the $G_{\sigma\pi\pi}(M_\sigma^2,M_\pi^2, M_\pi^2)$  is valid only when {\it  both $\sigma$ and $\pi$ are treated as pseudo NG bosons} in contrast to the low energy theorem Eq.(\ref{LETcoupling}) treating $\pi$ as a matter field, not as a pseudo-NG boson (or if we use the low energy theorem, we should regard $\pi$ as a matter field, not pseudo NG boson, i.e, put $M_\pi^2 \sim M_\rho^2$ as a typical matter field, in which case the width would also give a result roughly consistent with the reality, although $M_\pi^2$ is far from the reality).  %   

   \end{document}